\documentclass{article}
\usepackage[top=1.5in, bottom=1.5in, left=1.2in, right=1.2in]{geometry}
\usepackage{authblk}
\usepackage{amssymb, amsmath,framed}
\usepackage{graphicx}
\usepackage[round]{natbib}

\usepackage{bm}
\usepackage[colorlinks=true,linkcolor=blue,urlcolor=blue,citecolor=blue,anchorcolor=blue]{hyperref}
\expandafter\ifx\csname package@font\endcsname\relax\else
 \expandafter\expandafter
 \expandafter\usepackage
 \expandafter\expandafter
 \expandafter{\csname package@font\endcsname}
\fi
\def\bq{\begin{equation}}
\def\eq{\end{equation}}
\def\bqy{\begin{eqnarray}}
\def\eqy{\end{eqnarray}}

\makeatother

\begin{document}
\title{Implications of tides for life on exoplanets}

\author{Manasvi Lingam
 \thanks{Electronic address: \texttt{manasvi.lingam@cfa.harvard.edu}}}
\affil{Harvard-Smithsonian Center for Astrophysics, 60 Garden St, Cambridge, MA 02138, USA}
\affil{John A. Paulson School of Engineering and Applied Sciences, Harvard University, 29 Oxford St, Cambridge, MA 02138, USA}
\author{Abraham Loeb
 \thanks{Electronic address: \texttt{aloeb@cfa.harvard.edu}}}
\affil{Harvard-Smithsonian Center for Astrophysics, 60 Garden St, Cambridge, MA 02138, USA}

\date{}

\maketitle

\begin{abstract}
As evident from the nearby examples of Proxima Centauri and TRAPPIST-1, Earth-sized planets in the habitable zone of low-mass stars are common. Here, we focus on such planetary systems and argue that their (oceanic) tides could be more prominent due to stronger tidal forces.  We identify the conditions under which tides may exert a significant positive influence on biotic processes including abiogenesis, biological rhythms, nutrient upwelling and stimulating photosynthesis. We conclude our analysis with the identification of large-scale algal blooms as potential temporal biosignatures in reflectance light curves that can arise indirectly as a consequence of strong tidal forces.
\end{abstract}


\section{Introduction}
With rapid advances in the discovery of exoplanets over the past two decades \citep{WF15}, it has now become clear that there are a large number of Earth-sized planets in the habitable zone (HZ) - the region theoretically capable of supporting liquid water. In our Galaxy, it has been estimated that there may exist $\sim 10^{10}$ planets in the HZ of their host stars \citep{KKRH14,DC15}. Most observations to date have focused on low-mass stars, especially M-dwarfs, as they are more numerous, and any habitable planets orbiting them are easier to detect and analyze.

The program to detect planets in the HZ of M-dwarfs received two major boosts over the past year. The first was the discovery of Proxima b - the nearest exoplanet to the Earth at a distance of $1.3$ pc \citep{AE16}. The second was the discovery of at least seven terrestrial and temperate planets transiting the star TRAPPIST-1 at a distance of $12$ pc \citep{Gill16,Gill17}. The TRAPPIST-1 planetary system possesses several unique features such as mean-motion resonances \citep{Lug17} and, more importantly, the existence of potentially three planets in the HZ. The recent discovery of the rocky super-Earth LHS 1140b at a distance of $12.5$ pc \citep{DID17} also supports the notion that such planets are common.

As a result, the habitability of planets orbiting M-dwarfs has become a crucial question that has led to opposing viewpoints; for an overview of this rapidly evolving subject, the reader is referred to the reviews by \citet{Tart07,SKS07,SBJ16,Loeb}. Recent evidence, based on studies of the plasma environment of these planets, appears to suggest that the prospects for harbouring life are significantly lowered due to: (i) extreme space weather \citep{GDCAM,LM17}, and (ii) rapid atmospheric erosion by the stellar wind \citep{DLMC,Aira17,DHL17,LiLo17,Lin17}.

However, these results do not unequivocally imply that life cannot arise and evolve in M-dwarf planetary systems. For instance, the simulations performed by \citet{DJLAM} appear to suggest that the outer TRAPPIST-1 planets may retain their atmospheres over Gyr timescales. Another possibility is that life could be seeded by means of interplanetary panspermia, which has been argued to be greatly enhanced in the TRAPPIST-1 system \citep{LL17,Kri17}. Thus, these examples serve to clearly illustrate the fact that much work needs to be done before we can arrive at a proper understanding of habitability; although the latter concept is widely employed, it has also been misused at times \citep{MLJ17}.

In this paper, we broadly explore the potential implications of tides for habitability. Our premise stems from the fact that the tidal force scales inversely with the cube of the distance, thereby implying that stars less massive than the Sun could subject their planets in the HZ to strong tides. In Secs. \ref{SecHeat} and \ref{SecTid}, we estimate the extent of tidal heating and forces respectively with the TRAPPIST-1 system (and Proxima b) serving as an example. In Sec. \ref{SecBiol}, we explore the consequences of tides on key biological processes such as the origin of life, biological clocks and ocean mixing. We generalize our analysis to encompass other systems in Sec. \ref{SecOPS}, and delineate putative biosignatures arising from tidal effects in Sec. \ref{SecBioSig}. We conclude with a summary of our results in Sec. \ref{SecConc}.

\section{Tidal heating in the TRAPPIST-1 system and Proxima b}\label{SecHeat}
The role of tidal heating in planetary habitability has been thoroughly documented in the literature \citep{Barn17}. High tidal heating can result in continual volcanism (as on Io), thereby proving detrimental to life-as-we-know-it. On the other hand, it has been argued that it may also play an important role in initiating plate tectonics, and enhancing the prospects of habitability \citep{BJGR09}. This issue is all the more relevant for planets orbiting stars with masses $< 0.3 M_\odot$ since they are in real danger of undergoing rapid desiccation \citep{BMG13,DB15}; both Proxima b and the TRAPPIST-1 planets orbit stars that fulfill this criterion.

The most widely used estimate for tidal heating is the fixed $Q$ model \citep{PCR79,MuDe99}. The quantity of interest is the tidal heating rate $\dot{E}$, given by
\begin{equation} \label{TidalQ}
    \dot{E} = \frac{21}{2} \frac{k_2}{Q} \frac{R_p^5 n^5 e^2}{G}
\end{equation}
where $n = \sqrt{G M_\star/d^3}$ is the mean motion frequency, $R_p$ is the planetary radius, and $e$ is the eccentricity of the planetary orbit; $M_\star$ is the mass of the host star and $d$ is the semi-major axis of the planet. Here, $Q$ is the tidal dissipation factor and $k_2$ denotes the second Love number. This model is known to significantly underestimate the tidal heating \citep{MW07}, and determining $Q$ is challenging. Hence, viscoelastic models have been commonly employed as viable alternatives \citep{Ka64,SSRS}. The only difference is that the factor of $k_2/Q$ in (\ref{TidalQ}) is replaced by the imaginary part of the Love number:
\begin{equation} \label{Viscoel}
    \dot{E} = - \frac{21}{2} \mathrm{Im}\left(k_2\right) \frac{R_p^5 n^5 e^2}{G}.
\end{equation}
The exact expression for $\mathrm{Im}\left(k_2\right)$ depends on the specific viscoelastic model that is utilized; see Table 1 of \citet{HOCS09} for the corresponding formulae. The ``effective'' surface temperature of the planet $T_s$ arising solely from tidal heating can be estimated as
\begin{equation}
T_s = \left(\frac{\dot{E}}{4\pi R_p^2 \sigma}\right)^{1/4},
\end{equation}
where $\sigma$ is the Stefan-Boltzmann constant. Note that changing $n$ by a value of $\sim 10$ will result in a significant change in $ \dot{E}$ (by a factor of $\sim 10^5$) as seen from (\ref{Viscoel}).

We can rewrite (\ref{TidalQ}) in terms of normalized units,
\begin{eqnarray} \label{TidQNorm}
    \dot{E} &\approx& 50\, \mathrm{TW}\, \left(\frac{k_2}{0.3}\right)  \left(\frac{Q}{100}\right)^{-1} \nonumber \\
    && \quad \times \left(\frac{R_p}{R_\oplus}\right)^5 \left(\frac{n}{10^{-5}\,\mathrm{s}^{-1}}\right)^5 \left(\frac{e}{0.01}\right)^2,
\end{eqnarray}
where $R_\oplus$ is the Earth's radius, and the characteristic values of $n$ and $e$ have been chosen to be consistent with typical TRAPPIST-1 parameters \citep{WWBL}, while the fiducial values of $k_2$ and $Q$ have been chosen based on that of Earth \citep{DB15}. The typical value of $\dot{E}$ obtained from this simple estimate is (slightly) higher than the total internal heat generated by the Earth - this possibility was first noted by \citet{Lug17}, although a quantitative estimate was not provided therein.

However, the above formula should be seen as heuristic for a simple reason: the value of $Q$ can vary significantly depending on factors such as the planet's composition, orbital eccentricity, and orbital period. In fact, $Q$ can vary between $1$-$10^6$ for terrestrial exoplanets - see Table 2 of \citet{HH14} for further details. If we use (\ref{Viscoel}), the uncertainties in $Q$ are ``transferred'' to $\mathrm{Im}\left(k_2\right)$ as the latter can also vary by several orders of magnitude depending on the composition of the exoplanet. However, even taking the smallest value of $Q \sim 1$, tidal heating is not expected to exceed the insolation received by the TRAPPIST-1 planets; see also \citet{HOCS09} for a detailed discussion.

Before proceeding further, a few important points regarding tidal heating and $Q$ must be noted at this stage. Even a relatively small change in the value of $\dot{E}$ could lead to a significant change in the climate of distal planets. The planet can transition from being clement to either becoming ice-covered or subject to internally heated runaway greenhouse \citep{BMG13} depending on whether the flux is decreased or increased, respectively. Hence, an Earth-analog dominated by tidal heating would be unstable over time. Second, the value of $Q$ can change by an order of magnitude when the planet is subjected to tidal heating, and the role of self-organization should also be taken into consideration \citep{ZLDS}. Lastly, the value of $Q$ depends on a wide range of planetary factors, such as topography (the distribution of continents and oceans), plate tectonics, viscosity, temperature and rotation rate \citep{BCB10,Mel11}.

A similar calculation can be carried out for Proxima b, but a major difficulty stems from the fact that the eccentricity remains unknown since we only have an upper bound of $0.35$ at this time \citep{AE16}; recently, it was suggested that the most likely value may prove to be $e = 0.25$ \citep{Bro17}. As the tidal heating rate scales with $n^5$, we suggest that Proxima b's relatively wider orbit, leading to a lower value of $n$, may result in a lower value of $\dot{E}$ compared to most of the TRAPPIST-1 planets even if the eccentricity turns out to be higher.

It should be borne in mind that our discussion only pertains to the current epoch - at an earlier period, the planets could have been situated at greater distances from their host star and possessed a higher eccentricity.

\section{Tidal forces on the TRAPPIST-1 planets and Proxima b} \label{SecTid}
The tidal force per unit mass $a$ exerted on a given planet \citep{Cart99} can be written as
\begin{equation} \label{TidAcc}
    a \approx \frac{G M_\star R_p}{d^3},
\end{equation}
where we recall that $d$ is the star-planet distance and $M_\star$ is the stellar mass. Upon evaluating the ratio of the tidal acceleration exerted on the TRAPPIST-1 planets by TRAPPIST-1 (denoted by $a_T$) to that exerted on the Earth by the Sun (denoted by $a_E$), we find
\begin{equation} \label{TidARatio}
    \frac{a_T}{a_E} \approx 10^4 \left(\frac{R_p}{R_\oplus}\right) \left(\frac{d}{0.02\,\mathrm{AU}}\right)^{-3}\left(\frac{M_\star}{0.08\,M_\odot}\right),
\end{equation}
indicating that the tidal forces per unit mass are orders of magnitude higher for the TRAPPIST-1 planets compared to that experienced by the Earth. Note that we have normalized $d$, $R_p$ and $M_\star$ by the characteristic values for the TRAPPIST-1 system; here, $M_\odot$ denotes the mass of the Sun.

We now consider the tidal action of the star on a given TRAPPIST-1 planet assuming that it has a (relatively shallow) ocean that uniformly covers the surface. We wish to compute the rise in the water level as a result of the tidal forces. By adopting the tidal potential formalism \citep{Lamb32,Pugh96,But02}, the tidal elevation $H$ is estimated as
\begin{equation} \label{TideH}
H \approx \frac{3}{4} R_p \left(\frac{M_\star}{M_p}\right) \left(\frac{R_p}{d}\right)^3 \propto M_\star R_p^{0.3} d^{-3},
\end{equation}
where the second relation is valid only if one supposes that the composition of the planet is primarily rocky, as this enables us to use the mass-radius relationship $M_p \propto R_p^{3.7}$ for rocky planets \citep{VOCS,ZSJ16}. It is then evident from (\ref{TideH}) that the radius (or mass) dependence is very weak. When this formula is applied to the Earth-Sun system, a value of $H \approx 0.12$ m is obtained that is in excellent agreement with the more accurate estimate of $0.11$ m for the principal solar diurnal tide $S_2$ (see Table 17.2 of \citealt{Stew}). 

As before, let us express the amplitude of the ocean tides on the TRAPPIST-1 planets (denoted by $H_T$) in terms of the Earth-Sun value (denoted by $H_E$) by using (\ref{TideH}):
\begin{equation} \label{RelH}
\frac{H_T}{H_E} \approx 10^4 \left(\frac{R_p}{R_\oplus}\right)^4 \left(\frac{M_p}{M_\oplus}\right)^{-1} \left(\frac{d}{0.02\,\mathrm{AU}}\right)^{-3} \left(\frac{M_\star}{0.08\,M_\odot}\right),
\end{equation}
where $M_\oplus$ is the mass of the Earth. From (\ref{RelH}), it is apparent that tidal forces on the TRAPPIST-1 planets could result in considerable upsurges, conceivably upwards of $\mathcal{O}\left(10^2\right)$ m.

By using the tidal potential approach \citep{Cart99}, it is possible to classify the periods of the stellar tides. The diurnal (daily) and semidiurnal (twice-daily) tides are particularly noteworthy since they occur periodically with a period of $\tau_p$ and $\tau_p/2$ respectively, where $\tau_p$ is the planet's rotation period. Given that all of the TRAPPIST-1 planets orbit very close to their host star, it has been argued that they are likely to rotate synchronously \citep{Gill17} although higher-order spin-orbit resonances are theoretically possible, as noted for Proxima b \citep{Rib16}; they can, for instance, arise due to triaxial deformation \citep{ZL17}. Hence, it seems quite plausible that $\tau_p$ can be approximated by the orbital period for these planets.

At this stage, we wish to emphasize that the above estimates are based on the simplifying assumption of tidal equilibrium. In reality, tides are far removed from equilibrium, thereby necessitating accurate theoretical models that take into account a wide range of factors such as the topography, planetary rotation, and the elastic response to the tidal forces \citep{Pugh96,Stew}.

We observe that our estimates for $H_T$ and $\tau_p$ are only applicable to the \emph{present-day} TRAPPIST-1 system. It has been recently suggested, based on the orbital periods and the stability, that the TRAPPIST-1 system may have been characterized by convergent migration \citep{TRPM}. Consequently, the TRAPPIST-1 planets may have started their inward migration at distances $2$-$8$ times their present value \citep{UDHL}. In turn, this would result in a reduction of the ocean tide amplitudes by a factor of $\approx$ $10$-$500$. However, even if we consider the upper limit, the resultant value is still an order of magnitude higher than that of Earth; the same scalings are also applicable to the tidal acceleration computed in (\ref{TidARatio}).

Furthermore, it is quite plausible that the initial rotation rates of the TRAPPIST-1 planets may not be the same as their current values (which are also unknown). In our Solar system, it has been suggested that the Earth may have been a rapid rotator initially before it underwent a giant impact event that produced the moon, and subsequently lost angular momentum and was subject to despinning to attain its current value \citep{CS12,WT15}. Thus, the possibility that the TRAPPIST-1 planets may have had faster rotation rates in the past should not be overlooked. As a result, it does not automatically follow that their initial rotation periods were in a resonance with the corresponding orbital periods.

Lastly, our discussion focused on the TRAPPIST-1 planets, but most of the results are also applicable to Proxima b. The only difference is that the tidal forces and the elevation of the tidal bulge, (\ref{TidAcc}) and (\ref{TideH}) respectively, may be slightly lower for Proxima b compared to the TRAPPIST-1 system on account of: (i) the greater distance from the host star, and (ii) the higher mass of the host star ($0.12\, M_\odot$ instead of $0.08\, M_\odot$); note that $M_\odot$ denotes the mass of the Sun. Thus, our estimates for the aforementioned quantities may be lowered by about an order of magnitude for Proxima b. 

\section{On the biological consequences of tides} \label{SecBiol}
Below, we shall explore a select few (far-reaching) classes of biological phenomena that are potentially affected by tides, and thereby explore the ensuing implications for the TRAPPIST-1 system. 

\subsection{Tidal cycling and abiogenesis}
About a decade ago, \citet{Lathe04} proposed a theoretical model that posited tides as a major player in the origin of life (abiogenesis) on Earth. The central idea was that tides facilitated the chemical reactions analogous to the Polymerase Chain Reaction (PCR). The latter is characterized by a cyclical process oscillating between two different temperatures and driving DNA amplification in the presence of an appropriate polymerase enzyme \citep{Mull90,EGS91}. It was conjectured by \citet{Lathe04} that tides could drive cycles of flooding and drying that supplied the aforementioned temperature contrasts as well as variations in the salinity and prebiotic molecule concentrations necessary for polymerization and dissociation - more specifically, the polymerization operates during the drying/evaporation phase, whilst dissociation occurs during the dilution/flooding phase \citep{Lathe05}.

The model relied on two central assumptions: (A) the rotation rate of Earth at the time of life's emergence was faster, leading to semidiurnal tides with a lower periodicity ($\lesssim 6$ h), and (B) a closer Earth-Moon distance, which enabled the formation of tidal areas that ``\emph{extended several 100 km inland}'' \citep{Lathe04}. Both of these assumptions were rightly critiqued by \citet{VRD}, and Lathe revised some of his original predictions in subsequent publications \citep{Lathe05,Lathe06}. The tidal chain reaction (TCR) was further investigated by \citet{FVKS} who concluded that the mechanism was unlikely to be functional unless certain criteria, such as restriction ribosomes and complex (e.g. hairpin) RNA structures, were present. 

In addition, several studies have suggested that the Earth comprised mostly of oceans in the Hadean era \citep{IK10,AN12,Kam15}, and perhaps extending up to the late Archean era at $2.5$ Gya \citep{FCR08}; see, however, \citet{HSM08,Har09,HCD13,HCD17}. Thus, if continents emerged relatively late in Earth's history (post-abiogenesis), the role of tides in setting up wet-dry cycles would be rendered a moot point. Hence, on account of the above reasons, the TCR is not expected to constitute a viable mechanism insofar Earth is concerned. However, this fact ought not automatically imply that the relevance of the mechanism can be ruled out on all exoplanets in the HZ of their host stars.

We now turn our attention to (A) and (B), with the above caveats in mind. We begin by observing that condition (B) is likely to be valid in the TRAPPIST-1 system. As noted in Sec. \ref{SecTid}, even accounting for the original positions of the TRAPPIST-1 planets, the tidal forces and bulges are likely to have been orders of magnitude higher than on Earth. In turn, this might result in much larger areas being subjected to cycles of flooding and evaporation. The situation regarding $\tau_p$ is trickier since we cannot estimate the rotation rates at the time of abiogenesis. However, as these planets could have rotated faster in the past, condition (A) may also be satisfied.

The above statements can be viewed as part of a broader narrative in the context of abiogenesis. The importance of dry-wet cycles (for e.g. due to tides) in synthesizing prebiotic polymers has been analyzed from a thermodynamic perspective \citep{DD15,RD16}. Mineral surfaces have been documented to play a key role in several issues associated with the origin of life \citep{CSH12,RBD14}. They serve as a natural means of concentrating the dilute prebiotic ``soup'' and facilitate condensation, self-organization and stabilization \citep{FHLO,BCF05,CSH12}. Moreover, they play an important role in separating left- and right-handed molecules \citep{HS03}, thereby paving the way for homochirality - a salient feature of life. \citet{HS10} identified five factors of the Hadean geochemical environment that played a putative role in the origin of life: (i) chemical complexity, (ii) mineral-aqueous solution interfaces, (iii) gradients, (iv) fluxes, and (v) cycles \citep{Eig71}. Of these factors, \citet{HS10} have already pointed out that tides might have played a crucial role in regulating the factors (iv) and (v). 

We argue that, in planetary systems situated around low-mass stars, vigorous tides can result in a greater degree of flooding and evaporation, thereby enabling (ii), namely interfaces, to encompass larger areas. With regard to (iii), variations in salinity and other chemical concentrations, could arise through intensive tidal cycling provided that (A) and (B) are satisfied. These variations could potentially give rise to physicochemical gradients that play a crucial role in abiogenesis \citep{Spit13} such as, for instance, osmotic gradients that are necessary for chemiosmosis. The latter's importance stems from the fact that it plays a central role in energy metabolism, and has thus been invoked as a key player in the origin of life \citep{LAM10,SC13}. However, an important issue that affects planets around M-dwarfs is that they may end up being either desiccated or ocean-dominated \citep{TI15}. In general, neither of these two categories will be conducive to the emergence of life via tidal flooding-drying cycles, since the latter would typically be absent on such planets.  

\subsection{Tides and biological rhythms}
Most biological organisms on Earth have evolved internal timing mechanisms, often dubbed as biological ``clocks''. These clocks are found in some \emph{Prokaryota} \citep{DVN03}, and most members of \emph{Eukaryota} especially those belonging to the well-known kingdoms \emph{Plantae}, \emph{Fungi} and \emph{Animalia} \citep{Sog94}, thereby proving to be ubiquitous in complex organisms. Among these biological clocks, perhaps the best known are the circadian clocks that display a time period of approximately 24 h \citep{MSF,RMR}. 

The molecular bases of circadian clocks have been thoroughly explored in modern times \citep{Dun99,YK01}, with a particular emphasis on the central and peripheral mammalian circadian clocks \citep{RW02,DSA,MGT12}; the primary clock in \emph{Mammalia} is located in the suprachiasmatic nuclei (SCN) that resides in the anterior hypothalamus. It has been argued that circadian clocks have originated independently at least in two instances \citep{BPCE,Ros09}. If this hypothesis proves to be correct, circadian clocks would constitute an example of evolutionary convergence \citep{Mor03}. Circadian clocks play an important role in several biological contexts \citep{NS17}, for e.g. endowing fitness advantage \citep{OA98,Mic03,Shar03}, maintaining metabolic homeostasis \citep{FL12,MRP13}, and regulating ecologically significant processes such as reproduction and migration in animals \citep{Morg04}.

On Earth, the circadian clock of most mammals, and other organisms, is regulated by the light-dark cycle \citep{Pitt93,JES08,Al12}. However, we note that the circadian clock does exist even in mammals that are incapable of visual perception \citep{BDM} - for e.g. members of the genus \emph{Spalax} - and other factors such as temperature can also play a role in resetting the circadian clock \citep{Mro96,Dun99}. When we turn our attention to synchronously rotating planets around M-dwarfs, we are confronted with an immediate difference, namely, the possible absence of the light-dark cycle. 

This raises an important question as to how the ``daily'' rhythms and biological clock can be regulated on generic exoplanets in the HZ of M- and K-dwarfs, where the tidal forces are expected to be stronger (see Sec. \ref{SSecOtherS} for further details). Here, we reiterate that these planets need not have been tidally locked in the past, and may still be subject to librations in the current epoch. Moreover, asynchronous rotation may arise due to a multitude of reasons including triaxial deformation \citep{ZL17}, semiliquid interiors \citep{Mak15} and the existence of a sufficiently massive atmosphere \citep{LWMM15}. Below, we explore the hypothesis that ocean tides might play a significant role in governing the biological clocks of aquatic and semi-aquatic species.

Our proposal is motivated, and partly justified, by fairly extensive empirical evidence that has clearly demonstrated the role of the tidal cycle in regulating animal behavioral patterns on Earth \citep{Neu81,Gib92,Palm00,Bar05,WZ08,Tess11}. It seems likely that tidal clocks had an independent origin from their circadian counterparts, based on evidence from the crustacean \emph{Eurydice pulchra} \citep{Zha13}. We observe that the following phenomena have been observed to occur on a periodic basis: (i) variations in activity, especially in crabs belonging to the genera \emph{Uca} and \emph{Carcinus} respectively \citep{Palm73,LT07}, (ii) locomotion rhythms in aquatic animals, for e.g. in \emph{Limulus polyphemus} and \emph{Carcinus Maenas} \citep{Nay96,CKW}, (iii) egg-hatching and larval release activities \citep{Neu81,Thur04}, (iv) differences in oxygen consumption \citep{Palm95}, and (v) intertidal migration and selective tidal-stream transport in decapods and fishes \citep{Gib03}. In most of these cases, the rhythms recur over a period of $\tau_p/2$ or $\tau_p$, indicating that they are associated with semidiurnal (circatidal) or diurnal (circadian) tides respectively.

In this context, we also observe that the earliest concrete evidence of land animals on Earth dates back to the Silurian-Ordovician period \citep{JSE90,JBS94,Pis04}, although arthropod tracks have been dated to the Cambrian epoch \citep{MCD02,RSDP13}. Most fossils from this period are from the subphyla \emph{Myriapoda} (e.g. centipedes and millipedes) and \emph{Chelicerata} (e.g. arachnids and horseshoe crabs) that fall under the domain of \emph{Arthropoda} \citep{KN17}. As many aspects of the life cycle of present-day crabs, which are also arthropods, are governed by tides, we therefore can speculate that the first land-dwelling animals may also have been regulated by circatidal (and circalunar) rhythms to some degree \citep{Nay15}. If this hypothesis is correct, tides would have played an important role in the evolution of terrestrial life on Earth \citep{Bar05}, since animals play a notable role in nutrient cycling, regulating oxygen levels and ``engineering'' the biosphere \citep{Butt11,LRP14,Ver17}.

As we will see in Sec. \ref{SSecOtherS}, the tidal bulges for planets around M- and K-dwarfs are much more significant than on Earth. Hence, it stands to reason that the influence of oceanic tides on these planets (if present) on ecology and evolution would be more profound. Especially in the absence of the light-dark resetting mechanism, the tidal cycle may constitute an important biological clock, and thereby regulate biological properties from the ecosystem to the cellular level. However, an important caveat worth reiterating is that the tidal cycle is likely to affect aquatic and amphibious organisms to a considerable degree, but its influence on land-based lifeforms may be minimal.

\subsection{Turbulent mixing in oceans} \label{SSecMix}
It is well known that, in the absence of turbulent mixing, oceans on Earth would become stagnant and be characterized by very weak (convective) circulation \citep{FW09}. The energy required for ocean mixing is provided primarily by tides and winds \citep{WF04}. In particular, several studies have concluded that tides provide $\sim 50\%$ of the energy required for sustaining large-scale thermohaline ocean circulation \citep{MW98,ER00,ER01}. The source of this energy is tidal dissipation in the ocean interior partly due to internal wave generation and breaking by interactions with topographic gradients \citep{JSL01,GK07}. In this context, we point out that energetic internal waves can be excited at the semidiurnal tidal frequency \citep{IWK08}.

As noted in \citet{MW98}, most of the tidal dissipation arising from the Moon (and the Sun) is dissipated in the ocean, with only a small fraction being dissipated in the atmosphere and planetary interior. We now turn our attention to (\ref{TidQNorm}), from which we can see that the characteristic value of tidal energy dissipation is $\mathcal{O}(10)$-$\mathcal{O}(100)$ TW for planets around M-dwarfs. This is higher than the total tidal dissipation arising from the Moon by about $1$-$2$ orders of magnitude. Hence, if the energy budget for ocean mixing is similar to that of Earth ($\sim 2$ TW), it would appear as though tidal dissipation could play a dominant role in providing the mechanical energy for meridional overturning circulation (MOC) in planets orbiting low-mass stars.

However, the physical reality is not quite so simple. As noted earlier, the topography plays an important role in determining the generation and breaking of internal waves. From a theoretical standpoint, this is seen from the relevance of the steepness parameter, i.e. the topographic slope divided by the vertical length scale of these internal waves \citep{SLG02}. Hence, it automatically follows that the existence of plate tectonics, which plays a clear role in shaping bathymetric hetereogeneity \citep{Con89}, would constitute an important factor. The factors influencing plate tectonics on exoplanets are many, and this subject has a long and much-contested history which we will not address here. Initial studies were directed towards understanding whether plate tectonics is more likely or less likely on super-Earths \citep{VOCS07,ONC07}, although subsequent studies have pointed out the complex dependence of plate tectonics on the dynamical evolution of planetary properties \citep{WL12,SS16}.

Even if we assume that the requisite bathymetric heterogeneity exists such that it facilitates vertical transport, the strength of the mixing still remains unknown. The latter is quantified via the vertical diffusivity, which is known to depend on the depth \citep{BL79}, the stratification \citep{Garg84}, the topographical roughness \citep{PTLS97} and several other quantities \citep{Jay09}. The mixing will also be affected by the temperature profile through the ocean \citep{WF04}. It is widely known that vertical mixing is lowest in the relatively stable thermocline, and is enhanced at greater depths \citep{Gre87}. As M-dwarfs on the lower end of the mass spectrum will radiate primarily in the near-infrared, the stable stratification of the upper levels can be reduced since the radiation is strongly absorbed in this range \citep{CP51,KTJ07} and will therefore display a lesser degree of penetration. 

A potential ramification of the above point is that the stable stratification would be more pronounced on the day side than the night side, and simulations suggest that this is indeed the case \citep{DG17}, when other factors (e.g. salinity) are not included. Thus, as a consequence, it could be possible for strong mixing to occur closer to the surface on such exoplanets compared to Earth. A possible downside is that the euphotic zone would be shallower for the above reasons on planets that mostly receive near-IR radiation. This could, however, be partly offset by the fact that tidally locked planets are typically expected to have stronger winds than Earth \citep{Ed11,LingMa}. Thus, on such exoplanets, wind-driven circulation \citep{Alf01} is capable of giving rise to greater ocean mixing.

The importance of the above discussion stems from the fact that ocean mixing, which is driven primarily by tidal forces and winds, promotes the upwelling of nutrients to the euphotic zone \citep{SE84,HPM85,MG03,Scha10}. In turn, the enhanced availability of nutrients via ocean mixing will probably have a wide array of consequences for marine ecosystems \citep{WH88,SRB09}, biogeochemistry \citep{Sar06} and climate \citep{WNG06,DBI06}. Naturally, it goes without saying that the availability of nutrients is a necessity for the upwelling of nutrients to take place; the former factor will depend on several aspects, such as the composition of the mantle \citep{SBH13}, for e.g. whether it is predominantly ice/silicate.

\subsection{Other biological implications of tides} \label{SSecBiolT}
In addition to the above processes, tides play a major role in regulating coastal erosion, sediment movement, tidal currents, tidal mixing fronts, and strength of tidal streams.\footnote{The action of tides on seabeds in coastal or open ocean regions also leads to erosion \citep{CPO02} and changing chemical composition. Both of these factors have been documented to affect oceanic biogeochemical cycles over time \citep{Sar06,Sant08}; in particular, tidal effects on deep ocean ecosystems are likely to be more prominent \citep{Orc11,EBC12}.} All of these factors result in important biological long-term responses, and give rise to regions of high biological productivity in some instances \citep{LF87,LPQS}. High biodiversity in these habitats, for e.g. in intertidal zones \citep{RH99,DMS09}, is also typically accompanied by substantial environmental variability. 

The high biological productivity is a consequence of the fact that the tidal cycle maintains a continual supply of nutrients, and some of these environments (e.g. littoral zones) are also shallow enough to support photosynthesis \citep{Pugh96}. As both of these elements are fairly generic, it seems plausible that similar zones, characterized by a high degree of biodiversity and productivity, could be present on exoplanets. It is worth bearing in mind that photosynthesis on planets orbiting M-dwarfs, if prevalent, may produce spectral signatures in the infrared range instead of the conventional ``red edge'' \citep{KST07}.

Another important consequence of tidal forces (due to obliquity) is their capacity to generate large-amplitude Rossby waves in the open ocean \citep{Long68}. On Europa, it has been suggested that the kinetic energy associated with the flow is $\sim 10^{19}$ J, and that it can lead to tidal dissipation and heating of the oceans \citep{Ty08}. Even if the oceanic tidal heating contribution to the overall thermal budget is sub-dominant \citep{CNG14}, the generation of large-amplitude Rossby waves is an important biological result in its own right. 

The importance of Rossby waves stems from the fact that they can potentially stimulate photosynthesis \citep{ML06,Levy08} through wave-induced upwelling \citep{YO01,Sak04} via a ``rototiller'' mechanism that uplifts nutrients into the euphotic layer \citep{Sieg01}; on the other hand, the observed chlorophyll signals have also been explained through horizontal advection by the waves against a background gradient \citep{Kill04}. Rossby waves may also cause the convergence of floating particles on the surface, thereby serving as a marine ``hay rake'' \citep{DVL03}, although the feasibility of this mechanism has been critiqued by \citet{Ki04}. Lastly, wave-induced action leads to evolutionary advantages and enables accelerated development, suggesting that organisms can evolve and grow more rapidly when waves are present \citep{TR03}.

Hitherto, we have discussed the effects of tides on biological processes, but it is worth emphasizing that this picture is a simplified one. In reality, a reciprocal coupling between tide-dominated environments (e.g. estuaries) and biota is likely to be prevalent \citep{VTB11,WT13}; the situation could be quite analogous to the mutual feedback effects involved in shaping the co-evolution of geomorphology and ecosystems \citep{Cor11}. For instance, tidal dissipation facilitates vertical mixing in the ocean (Sec. \ref{SSecMix}), which thereby enables the delivery of nutrients to the euphotic zone, and could therefore play a beneficial role in sustaining ocean ecosystems. In turn, the marine biosphere can drive ocean mixing (and circulation) at a level that is ostensibly comparable to tides and winds \citep{DBI06,KD09} although the evidence has been debated \citep{Viss07}. Such co-evolution, if existent, can be viewed as a variant of the Gaia hypothesis that envisions life and the planetary environment as a coupled, co-evolutionary system \citep{LM74,Len98}.

Here, it must be recalled that, as with all other hypotheses concerning biological processes, there exist a multitude of physical, geological and biochemical factors that remain partly or wholly unknown. Hence, our conclusions concerning the role of the tides on abiogenesis, ecology and evolution in the TRAPPIST-1 system, Proxima b and other exoplanets must be interpreted with due caution. For instance, if the amplitudes of the tides are extremely high, they may have a detrimental effect on these phenomena. More specifically, massive tides could cause widespread habitat destruction unless they are mitigated by ecosystem resilience; mangrove forests represent a striking example of the latter since they attenuate the damage caused by tsunamis \citep{Al08}.

From the standpoint of abiogenesis, however, large tides ought not prove to be a major problem. Planets around low-mass stars (like the TRAPPIST-1 system) may have formed at greater distances and migrated inwards \citep{WF15}. Thus, it is conceivable that the tides, at the time of abiogenesis, were a factor of $\sim 10-100$ lower than their final value, as seen from the discussion regarding TRAPPIST-1 in Sec. \ref{SecTid}. A corollary of this fact is that the tidal elevation would be $\mathcal{O}(1)$-$\mathcal{O}(10)$ m; although this range is on the higher side, we note that waves of this amplitude have been documented on Earth even in the Anthropocene \citep{DKM08,Bry14}.

Furthermore, the time of abiogenesis is not properly constrained on Earth due to a scarcity of paleontological evidence \citep{KBS16}, and it is known that the Earth-Moon distance was initially $\sim 4\, R_\oplus$ \citep{Can12} compared to the current distance of $\sim 60\,R_\oplus$. Thus, from (\ref{TideH}) the lunar tidal forces could have been stronger (when life originated) by a factor of $\lesssim 10$ if the corresponding lunar distance was about half its present-day value. Taken collectively, the discrepancy between tides on Earth and those on planets orbiting low-mass stars might be reduced at the time of abiogenesis, implying that the latter is only about an order of magnitude higher than the former. 

Lastly, we can draw some insights from the studies of tsunamis, while being cognizant of the fact that they are \emph{not} simple oceanic tides. Nonetheless, if one supposes that they can be modeled via a solitary wave approach \citep{CTY03} - see, however \citet{CJ06} - a few interesting conclusions can be drawn. For values of $\mathcal{H}/\mathcal{D} \lesssim 0.2$, where $\mathcal{H}$ is the wave height and $\mathcal{D}$ is the depth of the water, it was found that wave breaking and dissipation of energy occurred before the shoreline was reached \citep{LR02,MFS08}; in some cases, large tsunamis bypassed islands and collided behind them \citep{LC95}. 

Empirical evidence from the 2009 South Pacific tsunami (with a height of $14$ m) was inundated $\sim 250$ m onshore \citep{AG11}. The height is comparable to tides on M-dwarf exoplanets provided that abiogenesis occurred when they were migrating inwards; it is also equal to the tidal amplitudes for exoplanets orbiting K-dwarfs. Even if the tides had a higher amplitude than $\mathcal{O}(10)$ m, abiogenesis could still be initiated inland, close to the point of inundation, where the level of sediment deposition is not so great, i.e. the dry-wet cycles would operate in a habitat (e.g. hydrothermal pools) not situated near the coast \citep{DD15}.

Thus, taken collectively all of these results appear to suggest that large-amplitude oceanic tides might not pose as much of a threat as one would expect for the origin of life.

\section{Tidal effects in other astrophysical systems} \label{SecOPS}
We will briefly discuss other possibilities (and systems), apart from TRAPPIST-1 and Proxima Centauri, wherein tidal effects are likely to play an important role. 

\subsection{Tides caused by the TRAPPIST-1 planets}
Hitherto, when discussing the TRAPPIST-1 system, we have restricted ourselves to the tidal force exerted by the host star, as it constitutes the most dominant source. Now, we turn our attention to the additional forces exerted by one planet on another; we do not consider tidal effects due to (large) exomoons in the TRAPPIST-1 system since they are unlikely to survive dynamically \citep{Kane17}.

Suppose that we consider the tidal force exerted by planet II (with mass $M_2$) on planet I (with mass $M_1$ and radius $R_1$), and the two planets are separated by a distance $d_{12}$; for the sake of simplicity, we assume that the two planets are nearest neighbours. Along the lines of our treatment in Sec. \ref{SecTid}, we can express the tidal elevation (\ref{TideH}) in terms of the Earth-Sun value:
\begin{equation}
\frac{H_{12}}{H_E} \approx 3 \left(\frac{R_1}{R_\oplus}\right)^4 \left(\frac{M_2}{M_1}\right) \left(\frac{d_{12}}{0.01\,\mathrm{AU}}\right)^{-3},
\end{equation}
where we have normalized $d_{12}$ in terms of the characteristic distance of closest approach between the two neighbouring planets of the TRAPPIST-1 system. 

Since the tides caused by the Moon (on Earth) are about twice as large as those caused by the Sun, we are led to the following conclusion: the tidal forces exerted by a nearest neighbour pair of TRAPPIST-1 planets on each other are approximately equal to that exerted by the Moon on the Earth. As the magnitudes of the tidal bulges are similar to those present on Earth, most of our preceding discussion concerning terrestrial oceanic tides and their biological consequences may also prove to be applicable here. 

In addition, we note that the existence of mean-motion resonances \citep{Lug17} in the TRAPPIST-1 could enhance tidal effects by causing the respective amplitudes to ``add'' up or be subject to further modulation. A somewhat analogous situation is also seen on Earth when the Moon and Sun are collinear with the Earth, have zero declination, and at simultaneous perihelion (with respect to Earth) - this scenario results in maximum semidiurnal tides \citep{Pugh96}. Recently, \citet{Bal14} proposed that tidal modulation was responsible for the formation of a complex network of shallow pools on Earth which led to the development of chiridian limbs in tetrapodomorphs (enabling efficient motor control) during the Devonian period \citep{DSJF}.

As there will be \emph{six} planets exerting tidal forces on the seventh, and given the existence of orbital resonances, it seems reasonable to hypothesize that the tidal forces could be amplified (by a factor of a few) in certain instances. Generally speaking, we anticipate that the rich dynamical behaviour of the TRAPPIST-1 system \citep{Lug17,TRPM} will translate to equally diverse tidal modulations (with multiple periods) that may have interesting consequences for abiogenesis and the evolution of life on these planets in a manner similar to that of Earth.

\subsection{Looking beyond the TRAPPIST-1 system} \label{SSecOtherS}

We now interpret the results from Secs. \ref{SecHeat} and \ref{SecTid} in a more general setting. We hold the mass, radius, orbital eccentricity and effective temperature of the planet to be fixed (equalling that of Earth), since we are interested in determining how the properties of the host star influence the tidal forces; the planet can thus be interpreted as an Earth-analog, albeit in a superficial sense.

Along the lines of \citet{LiLo17}, we choose $L_\star \propto M_\star^3$ as an approximate mass-luminosity relationship. For fixed values of the planetary parameters, it is easy to show that $d \propto L_\star^{1/2}$, thereby leading us to $d \propto M_\star^{3/2}$. Thus, we are led to the relations:
\begin{eqnarray} \label{MDepPar}
\dot{E} &\propto& n^5 \propto M_\star^{-35/4}, \\
a &\propto& \frac{M_\star}{d^3} \propto M_\star^{-7/2}, \\
H &\propto& \frac{M_\star}{d^3} \propto M_\star^{-7/2}. 
\end{eqnarray}
Hence, it is clear that the tidal heating, force, and elevation experienced by the Earth-analog display a strong dependence on the mass of the host star. For instance, if we choose a host star that is one-half the Solar mass, it will result in tidal heating that is $\sim 400$ times higher than Earth. Moreover, the tidal force (per unit mass) and bulge experienced by this Earth-analog will be an order of magnitude higher than the Earth-Sun system. 

Thus, it is clear that most of our preceding discussion concerning the biological implications of tides are also applicable in general to planets in the HZ of M- and K-dwarfs. Even though the tidal forces are stronger on M-dwarfs, it does not necessarily mean that such stars are more suited for life to originate and evolve; there are a host of other factors that must be taken into consideration, for e.g. atmospheric losses and ultraviolet radiation \citep{SBJ16,Lin17,ManLo17,LiLo17,LL18,LiL18}, which can lower the prospects of life inhabiting M-dwarf planetary systems. 

Based on the above scalings, we argue that the role of tides are likely to be important when the primary object has a mass lower than the Sun provided that the eccentricity is not too low. Hence, our results are also applicable to planets around brown dwarfs, and Mars/Earth-sized moons orbiting Jovian planets. We observe that physical, but not biological, tidal effects in both these systems have been quite extensively investigated in recent times \citep{BH13,HB13,DT15,DHT17}.

\section{On the detection of biotic phenomena associated with tides}\label{SecBioSig}
The identification of viable biosignatures constitutes one of the most important endeavours in the arena of astrobiology \citep{Kal10,Po17,Sch17}, as it seeks to identify methods by which extraterrestrial life can be detected. Apart from the well-known paradigm of biosignature gases such as oxygen and methane, signatures for widespread features, like oceans \citep{WG08,Cow09} and vegetation \citep{FST01,STSF}, have also been identified. In addition, methods have been proposed for discerning microbial biospheres in extreme environments \citep{OGRC13,Heg15}, and other pigmented organisms \citep{SCM15}. Here, we focus on certain algal biosignatures that are intimately linked with the existence of oceanic currents and strong tides.

On Earth, harmful algal blooms (HABs), often referred to colloquially as ``red tides'', are a ubiquitous phenomenon \citep{Hall93,And94,AGB02,ACH12}. HABs are the manifestation of the explosive growth of algae in a wide range of aquatic environments. Most of the red tides in marine environments like ocean coastlines have been caused by dinoflagellates, of which the best known is the organism \emph{Karenia brevis} that has been responsible for several red tides observed in the Gulf coasts of Florida and Texas. The concentration of \emph{K. brevis} can amplify from $1$-$10^3$ cells/L (background) to levels reaching $10^6$-$10^7$ cells/L \citep{TS97,Pierce08}.

The exact relationship between HABs and eutrophication (nutrient enrichment of water bodies) has been extensively probed in recent times, since the latter can stimulate the formation of algal blooms. Although a wide range of anthropogenic and natural causes \citep{MB96} have been identified, \citet{And08} concluded that ``large-scale HABs along \emph{open coasts} appear to be unrelated to anthropogenic nutrients'', and these \emph{large-scale} HABs may rely upon nutrients supplied through upwellings or advection. In Sec. \ref{SSecBiolT}, we briefly delineated the role of tides in supplying nutrients through these avenues. Hence, planets with significant oceanic tides may potentially engender algal blooms that span a greater area, and recur more often. 

HABs on Earth can span a few square kilometers, although there appears to be no \emph{a priori} reason why they cannot be much larger on other planets. Spectral curves have been developed for algal blooms that accurately account for the effects of backscattering and absorption due to water, phytoplankton, and detritus \citep{CS85}, thereby enabling the identification of potentially unique spectral signatures of HABs \citep{Die06}. Several other methods have been identified for carrying out remote sensing observations of HABs \citep{SDK03,Hu05,Kut09} based on monitoring the likes of sea surface temperature, fluorescence data and satellite ocean colour imagery; matching the measured spectral reflectance with an ocean colour inversion model enables the determination of the algal bloom composition and concentration \citep{Stu03,BB14}. The detection of algal blooms can be viewed as a particular case of algal detection on exoplanets \citep{Sch17}, but there exists one crucial difference: as HABs are temporally \emph{transient}, one would anticipate changes in the spectral reflectance to occur over a relatively short timescale that might be quasi-periodic in nature. Thus, HABs fall under the category of \emph{temporal} surface biosignatures \citep{Sch17}.

Apart from the red tides discussed earlier, there has been a great deal of attention also devoted to ``green'' and ``golden'' seaweed tides \citep{SZ13}, as well as brown and black tides. The former, caused by the seaweed \emph{Ulva prolifera}, gave rise to presumably the world's largest macroalgal bloom ever recorded, spanning about $400$ square km \citep{Hu10}. The golden seaweed tides are caused by \emph{Sargassum natans} and \emph{Sargassum fluitans}, and have often been observed along the coastline of the Gulf of Mexico \citep{GK11}. Brown tides are algal blooms arising from the brown algae: \emph{Aureococcus anophagefferens} and \emph{Aureoumbra lagunensis} \citep{GS12}. In contrast, one of the causes for ``black tides'' are oil spills; on Earth, the latter can arise as a result of anthropogenic or natural (e.g. abyssal vents) phenomena. Hence, oil spills do not necessarily imply the existence of industrial pollution, although this possibility should be borne in mind. If they have been emerged due to anthropogenic reasons, oil spills would serve as \emph{technosignatures} signifying the presence of intelligent life.

The mechanisms proposed for green tides encompass high nutrient supply and uptake rates, as well as hydrodynamic seaward transport \citep{Liu13}. As most of these factors, at least in principle, could be enhanced due to the existence of stronger tides, the counterparts of red, green, golden and brown tides may constitute plausible biosignatures. However, we must add the caveat that these phenomena need not be caused exclusively by tidal forces; we have implicitly conjectured that tides play an indirect role in generating and sustaining them. 

Lastly, we have already indicated in Secs. \ref{SecTid} and \ref{SSecOtherS} that oceanic tides are significantly higher in some planetary systems, and that tidal forces can generate large-amplitude Rossby waves, conceivably even orders of magnitude higher than on Earth. Remote sensing measurements have made it possible to study ocean currents \citep{GBZ89}, Rossby waves \citep{CCC01}, and other ocean surface phenomena \citep{Mart14} on Earth. Hence, the direct detection of large-amplitude Rossby (or internal) waves could also be indicative of the existence of strong tidal forces, although they are not biosignatures by themselves. Moreover, we caution that Rossby waves can be excited due to a wide array of mechanisms not associated with tides \citep{Val17}, and should therefore not be perceived as unequivocal evidence in favor of tides.

For an overview of the prospects of detecting biosignatures from an observational standpoint, we refer the reader to \citet{Fuj17}. In Section 4.3.2, the authors concluded that high-contrast imaging could enable the detection of surface biosignatures such as the red edge of vegetation and other biological pigments. Based on our preceding discussion, algal blooms belong to this category as they are characterized by a number of unequivocal spectral features, for e.g. the chlorophyll fluorescence peak at $0.683$ $\mu$m. Although the Wide-Field Infrared Survey Telescope (WFIRST),\footnote{\url{https://wfirst.gsfc.nasa.gov/}} equipped with a starshade, is capable of achieving a contrast of $10^{-10}$, such phenomena are more likely to be detectable by post-2030 missions such as Habitable Exoplanet Imaging Mission (HabEx),\footnote{\url{https://www.jpl.nasa.gov/habex/}} and Large UV/Optical/Infrared Surveyor (LUVOIR).\footnote{\url{https://asd.gsfc.nasa.gov/luvoir/}}

However, an important point to note is that the detection of the above phenomena presupposes the existence of an ocean, and it is therefore necessary to take into account the effects of climate \citep{YCA13,FL14,Wolf17}, and water loss from atmospheres \citep{LB15,Bol17,DHL17}. Moreover, missions like LUVOIR would focus primarily on planets around K- and G-type stars \citep{DS17}, where the tidal forces on exoplanets are likely to be weaker compared to those orbiting M-dwarfs. Hence, future analyses of tide-mediated events should attempt to identify appropriate biosignatures that are detectable through transit transmission spectroscopy.

\section{Conclusion} \label{SecConc}
On Earth, the presence of (oceanic) tides has revealed their importance in the context of a wide range of biological phenomena. In this paper, we explored tidal effects on exoplanets and proposed how tides may affect, or regulate, the origin and evolution of life.

We began by estimating the degree of tidal heating experienced by the planets of the recently discovered TRAPPIST-1 system due to the host star. In light of the many uncertainties involved regarding the composition of these planets, a precise estimate is not possible. However, despite the accompanying variability, we showed that the tidal heating on most of these planets could exceed the total internal heat generated by the Earth, in agreement with previous results \citep{Lug17}. 

We continued by further estimating the tidal acceleration and elevation (height of the tidal bulge) for these planets, and showed that they are orders of magnitude higher than that of Earth. Our results are also broadly applicable to other exoplanets orbiting low-mass stars such as Proxima b. We also computed the tidal forces exerted by nearest neighbours of the TRAPPIST-1 system on each other, and concluded that the tidal effects are comparable to the lunar tides on Earth.

As exoplanets closer to their host star (compared to the Earth-Sun system) can experience tides with a larger amplitude, we explored the role of tides in influencing many aspects of life-as-we-know-it. We have delineated some of the most plausible and salient phenomena below:
\begin{itemize}
    \item Tides have been hypothesized to be responsible for cycles of flooding and evaporation (wetting and drying) that generate temperature and concentration gradients necessary for the polymerization and replication of self-replicating molecules \citep{Lathe04}. Hence, tides may be of some importance in enabling the origin of life on certain planets.
    \item Tides could play an important role in: (i) fluid circulation and mixing, (ii) applying selection pressure to prebiotic molecules through cycling, (iii) setting up chemical gradients, and (iv) exposing mineral-solvent interfaces \citep{HS10}. Each of these factors might have been potentially important for life to originate.
    \item On planets where the light-dark cycle is absent, the tidal cycle may function as an effective biological ``clock''; the latter has been shown to confer enhanced evolutionary fitness and metabolic homeostasis on Earth. Circatidal rhythms have been observed in several animals on Earth (e.g. fishes and crabs) and are likely to be more important on planets with stronger tides.
    \item Tidal forces are known to be responsible, through a variety of mechanisms, for facilitating turbulent vertical mixing in the ocean \citep{FW09}. In turn, vertical mixing can, under the right conditions, induce nutrient upwelling and convergence, and thereby stimulate algal photosynthesis \citep{ML06}.
    \item The coupling between tide-dominated environments and ecosystems is likely to result in their co-evolution via mutual feedback mechanisms \citep{CZVM13}.
\end{itemize}
We theorized that tides, due to a number of reasons, can lead to enhanced hydrodynamic transport, for e.g. by means of turbulent vertical mixing of the ocean and generating planetary waves. This factor has been argued to be one of the putative causes behind harmful algal blooms (HABs) that represent large-scale agglomerations of algae; on Earth, they have encompassed areas exceeding $100$ square km. HABs are characterized by a distinctive colour (e.g. red, green, brown) and spectral signatures, and thus may constitute genuine \emph{temporal} biosignatures in surface reflectance light curves. Although their scope of detection lies beyond that of JWST, future telescopes such as WFIRST, HabEx and LUVOIR may be capable of detecting these phenomena.

Owing to the ubiquity of tides on Earth, their significance has, in some cases, been overlooked or insufficiently appreciated. In the case of exoplanets orbiting closer to their host star, the tides thus generated may be much stronger than on Earth.\footnote{Here, it must be recognized that stronger tides could also result from the existence of large moons, or when multiple exoplanets are closely clustered together. For both these scenarios, the general conclusions presented in the paper are likely to be valid.} Hence, if one extrapolates tide-mediated phenomena to such exoplanets based on our understanding of life-as-we-know-it, the consequences merit further study. As we argue in our paper, this is because of the fact that, in certain instances, tides can: (i) play a potentially significant role in facilitating abiogenesis, (ii) serve as an influential biological clock in some habitats, and (iii) lead to enhanced hydrodynamic transport and boost nutrient distribution and photosynthesis. Thus, future investigations of habitability, especially those undertaken from a biological perspective, should duly take tidal effects into consideration.

\subsection*{Acknowledgements}
We thank John Barrow, Adam Frank, Norm Sleep, and the referees for their helpful comments concerning the paper. This work was partly supported by grants from the Breakthrough Prize Foundation for the Starshot Initiative and Harvard University's Faculty of Arts and Sciences, and by the Institute for Theory and Computation (ITC) at Harvard University. 


\end{document}